\documentclass[submission, Proceedings]{SciPost}

\begin{document}

\begin{center}{\Large \textbf{Tau Polarimetry in B Meson Decays
}}\end{center}

% TODO: write the author list here. Use initials + surname format.
% Separate subsequent authors by a comma, omit comma at the end of the list.
% Mark the corresponding author with a superscript *.
\begin{center}
R. Alonso\textsuperscript{1},
J. Martin Camalich\textsuperscript{2,3},
S. Westhoff\textsuperscript{4*}
\end{center}

\begin{center}
{\bf 1} Kavli Institute for the Physics and Mathematics of the Universe (WPI)
University of Tokyo, Kashiwa, Chiba, 277-8583 Japan
\\
{\bf 2} Instituto de Astrof\'isica de Canarias, C/ V\'ia L\'actea, s/n E38205 - La Laguna (Tenerife), Spain
\\
{\bf 3} Universidad de La Laguna, Departamento de Astrof\'isica, La Laguna, Tenerife, Spain
\\
{\bf 4} Institute for Theoretical Physics, Heidelberg University, D-69120 Heidelberg, Germany
\\
* westhoff@thphys.uni-heidelberg.de
\end{center}

\begin{center}
\today
\end{center}

\definecolor{palegray}{gray}{0.95}
\begin{center}
\colorbox{palegray}{
  \begin{tabular}{rr}
  \begin{minipage}{0.05\textwidth}
    \includegraphics[width=8mm]{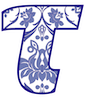}
  \end{minipage}
  &
  \begin{minipage}{0.82\textwidth}
    \begin{center}
    {\it Proceedings for the 15th International Workshop on Tau Lepton Physics,}\\
    {\it Amsterdam, The Netherlands, 24-28 September 2018} \\
    \href{https://scipost.org/SciPostPhysProc.1}{\small \sf scipost.org/SciPostPhysProc.Tau2018}\\
    \end{center}
  \end{minipage}
\end{tabular}
}
\end{center}

% For convenience during refereeing: line numbers
%\linenumbers

\section*{Abstract}
{\bf
This article summarizes recent developments in $B\to D^{(\ast)}\tau\nu$ decays. We explain how to extract the tau lepton's production properties from the kinematics of its decay products. The focus is on hadronic tau decays, which are most sensitive to the tau polarizations. We present new results for effects of new physics in tau polarization observables and quantify the observation prospects at BELLE II.}

\section{Introduction}
\label{sec:intro}
Semi-leptonic $B$ meson decays with tau leptons are important probes of lepton universality in weak interactions. By building ratios of decay rates, very clean observables can be predicted that depend very little on hadronic uncertainties. The detection of tau leptons is challenging, but recent progress at BELLE and LHCb shows that measurements beyond total decay rates are possible. This makes $B\to D\tau\nu$ and $B\to D^\ast\tau\nu$ decays a key area of study at the new BELLE II experiment.

On the theory side, the precision is limited by the hadronic form factors governing $B\to D$ and $B\to D^\ast$ transitions. Very recently, lattice calculations of vector and scalar form factors beyond the limit of zero $D^{(\ast)}$ recoil became available.~\footnote{See talks by Chris Monahan and Alejandro Vaquero at CKM 2018.} Combined with shape predictions based on heavy quark effective theory and QCD sum rules, they allow us to predict differential distributions in $B\to D^{(\ast)}\tau\nu$ to percent precision. This provides us with a solid basis to investigate semi-tauonic $B$ decays over the entire phase-space region.

Among the differential observables, lepton and meson polarizations are particularly interesting, because they are direct probes of the underlying interaction. Two recent analyses have pioneered polarization measurements in semi-tauonic $B$ decays. The BELLE collaboration has extracted the longitudinal tau polarization in $B\to D^{\ast}\tau\nu$,~\cite{Abdesselam:2016xqt,Hirose:2016wfn}
\begin{align}
P_L(\tau) = - 0.38 \pm 0.51\,(\text{stat})^{+0.21}_{-0.16}\,(\text{syst}).
\end{align}
The uncertainty is large, but statistics-dominated, which will be overcome at BELLE II. In the same process, BELLE has measured the fraction of longitudinally polarized $D^\ast$ mesons,\footnote{This result was presented by Karol Adamczyk at CKM 2018.}
\begin{align}
F_L(D^\ast) = \frac{\Gamma(D_L^\ast)}{\Gamma(D_L^\ast) + \Gamma(D_T^\ast)} = 0.60 \pm 0.08\,(\text{stat}) \pm 0.04\,(\text{syst}).
\end{align}
The remarkable precision of this measurement suggests that combined analyses of tau and $D^{\ast}$ observables have a good potential to probe the underlying production process.

In addition to these developments, observed deviations in the branching ratios of $B\to D^{(\ast)}\tau\nu$ decays, widely referred to as the $R_{D^{(\ast)}}$ puzzle, have inspired searches for new particles with non-universal lepton interactions. Regardless of the motivation to explore $B\to D^{(\ast)}\tau\nu$ decays, however, any information about these processes is imprinted on the tau and $D^{(\ast)}$ decay products. The presence of one or more neutrinos in the final state prevents us from fully reconstructing the decay kinematics. In order to gain maximal information on the production process, it is thus necessary to define observables in terms of these decay products.

Recent analyses of $B\to D^{(\ast)}\tau\nu$ differential distributions have pursued two methods: a numerical approach based on Monte-Carlo simulations~\cite{Ligeti:2016npd}, and an analytic approach based on a fully analytic phase-space integration~\cite{Becirevic:2016hea,Alonso:2016gym,Ivanov:2017mrj,Alonso:2017ktd}. While numerical simulations can be directly used at an experiment, analytical predictions make the relations between observables and features of the underlying process transparent. In this article, we pursue the analytic approach to extract tau polarizations and a forward-backward asymmetry in tau production from final-state kinematics. The discussion will be focused on $B\to D \tau \nu$; an extension to $B\to D^{\ast} \tau \nu$ is possible by pursuing the same strategy.

Beyond the standard model, tau polarimetry is an important tool to detect and disentangle possible new physics effects. We probe the sensitivity of tau polarization observables to new scalar, vector and tensor currents contributing to $B\to D \tau \nu$ decays in Section~\ref{sec:np}.

\section{Tau properties in $B\to D \tau \nu$ decays}
\begin{figure}[t]
\centering
\includegraphics[height=1.5in]{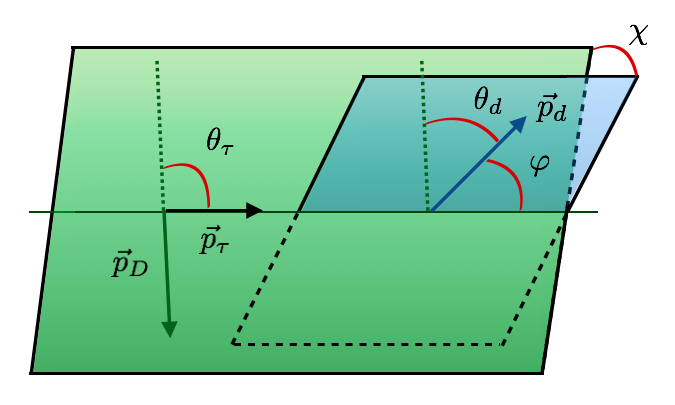}
\caption{Kinematics of the decay chain $B\to D \nu [\tau \to d \nu(\bar{\nu})]$, where $d=\{\pi,\rho,\ell\}$. The momenta of the $D$ meson and the tau lepton span the plane in $B\to D\tau\nu$ decays (in green). The momenta of the tau lepton and its visible decay product $d$ span a second plane in the subsequent decay $\tau\to d\nu(\bar{\nu})$ (in blue).}
\label{fig:kinematics}
\end{figure}

We start by briefly reviewing the observable tau properties in $B\to D \tau \nu$ decays. For more details, we refer you to our Ref.~\cite{Alonso:2017ktd}. The differential decay rate for producing a tau lepton polarized along a certain direction $\hat{s}$ is given by
\begin{align}\label{eq:pol}
d\Gamma(\hat s)=\frac{d\Gamma}{2}\Big[1 + \left(dP_L\,\hat e_\tau+dP_{\perp}\,\hat  e_\perp+dP_T\,\hat e_T \right)\cdot \hat s\Big].
\end{align}
We choose our coordinate system $\{\hat e_\tau, \hat e_\perp,\hat e_T\}$ as
\begin{align}\label{eq:coords}
\hat e_\tau&=\frac{\vec p_\tau}{|\vec p_\tau|},\qquad \hat e_T =\frac{\vec p_D\times\vec p_\tau}{|\vec p_D\times\vec p_\tau|}, \qquad \hat e_\perp =\hat e_T\times \hat e_\tau,
\end{align}
where $\vec p_{\tau}$ and $\vec p_D$ are the momenta of the $\tau$ lepton and the $D$ meson defined in the rest frame of the tau-neutrino pair. The decay kinematics are illustrated in Fig.~\ref{fig:kinematics}. The longitudinal polarization $P_L$ thus points in the direction of the tau momentum, the perpendicular polarization $P_\perp$ lies in the $D-\tau$ plane perpendicular to the tau momentum, and the transverse polarization points orthogonal to the $D-\tau$ plane. Here we will focus on the longitudinal and perpendicular polarizations. Observing the transverse polarization requires information beyond the currently accessible decay kinematics.

%---------------------------------------------------------------------------
%---------------------------------------------------------------------------

The unpolarized rate for $B\to D\tau\nu$ depends on the angle between the tau and $D$ momenta, $\cos\theta_\tau$. We define the tau forward-backward asymmetry with respect to the $D$ meson momentum as
\begin{align}
\label{eq:dAFBdq2} \frac{d\Gamma}{dq^2}A_{\tau}(q^2) & = \int_0^1 d\cos\theta_\tau \frac{d^2\Gamma}{dq^2 d\cos\theta_\tau} - \int_{-1}^0 d\cos\theta_\tau \frac{d^2\Gamma}{dq^2 d\cos\theta_\tau},
\end{align}
where $q$ is the four-momentum of the tau-neutrino pair.

% Integrated over the full phase space, the tau polarizations and forward-backward asymmetry in the standard model are
% \begin{align}\label{eq:properties}
% P_L = 0.34(3),\quad P_\perp = -0.839(7),\quad A_\tau = -0.359(3).
% \end{align}
% The uncertainties are due to our limited knowledge of the relevant hadronic form factors. In Fig.~\ref{fig:properties}, left, we show these tau properties as functions of the momentum transfer $q^2$.

%---------------------------------------------------------------------------
\section{Observables from tau decay products}
The tau properties above are not directly observable, but need to be reconstructed from the tau decay products. We consider the three most frequent tau decay modes, $\tau\to \ell\bar{\nu}\nu$ $(\ell = e,\mu)$, $\tau\to\pi\nu$, and $\tau\to\rho\nu$. The hadronic decays $\tau\to\pi\nu$, and $\tau\to\rho\nu$ have a higher analyzing power than $\tau\to \ell\bar{\nu}\nu$, because only one neutrino is missed. The momenta of the visible decay product $d=\{\ell,\pi,\rho\}$ and the tau momentum span a second plane, see Fig.~\ref{fig:kinematics}. Due to the missed neutrino momentum, the relative orientation between the $D-\tau$ plane and the $\tau-d$ plane is not observable.

The $D$ meson and the decay product $d$ are the only visible particles in the final state. The maximal kinematic information in the full decay process $B\to D\nu [\tau \to d \nu(\bar{\nu})]$ can be expressed in terms of three variables. We choose them as the momentum transfer, $q^2$, the energy of the decay particle, $E_d$, and the angle between the $D$ and $d$ momenta, $\cos\theta_d$, all defined in the tau-neutrino rest frame. The differential decay rate can be written as
\begin{align}\label{eq:triple-rate}
\frac{d^3\Gamma_d}{dq^2\,dE_{d}\,d\cos\theta_{d}} = \mathcal{B}_{d}\frac{\mathcal{N}}{2m_{\tau}}\big[I_0(q^2,E_{d}) + I_1(q^2,E_{d})\cos\theta_d + I_2(q^2,E_{d})\cos^2\theta_{d} \big],
\end{align}
where $\mathcal{B}_{d}$ is the branching ratio for $\tau \to d \nu(\bar{\nu})$, $\mathcal{N}$ is a normalization factor, and the kinematic functions $I_i$ are defined in Ref.~\cite{Alonso:2016gym}. After integrating over the angle $\cos\theta_d$, we obtain the double-differential rate
\begin{align}
\frac{d^2\Gamma_d}{dq^2\,ds_{d}} = \mathcal{B}_{d}\frac{d\Gamma}{dq^2}\Big[f_0^d(q^2,s_d) + f_L^d(q^2,s_d) P_L(q^2)\Big],\qquad s_d = E_d/\sqrt{q^2}.
\end{align}
The energy distribution of the decay particle $d$ is thus sensitive to the longitudinal tau polarization. Notice that for $\tau\to\pi\nu$ and $\tau\to\rho\nu$ the function $f_0^d$ does not depend on $s_d$, so that the energy distribution is maximally sensitive to $P_L$.

\begin{figure}[t]
\centering
% \hspace*{-0.7cm}\includegraphics[height=2.1in]{fig2a.pdf}\hspace*{0.2cm} 
\includegraphics[height=2.03in]{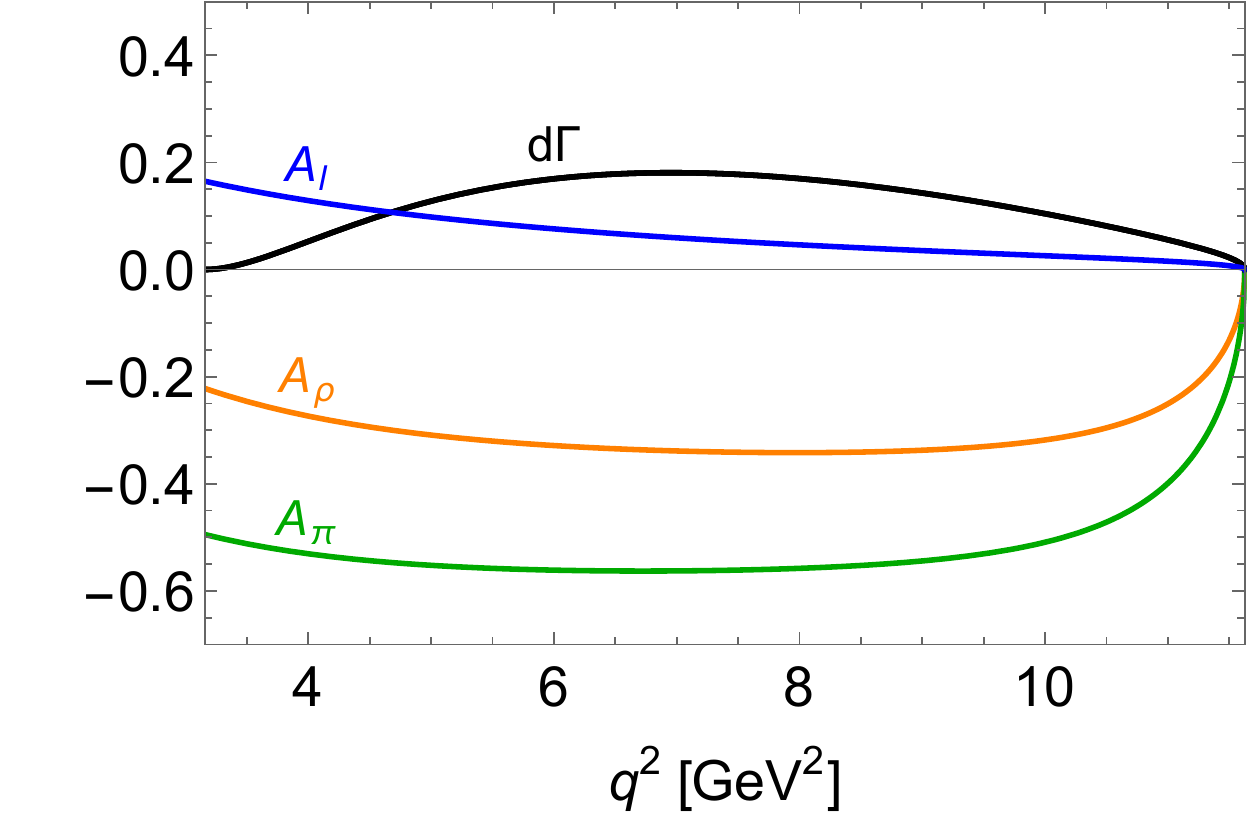}
\caption{
% Left: Tau polarizations $P_L,\,P_\perp$ and forward-backward asymmetry $A_\tau$ in $B\to D \tau\nu$ as a function of the momentum transfer $q^2$. Right: 
Decay rate $d\Gamma$ and angular asymmetries $A_d$ of final-state particles $d=\pi,\rho,\ell$.}
\label{fig:properties}
\end{figure}

Complementary to $d^2\Gamma_d/dq^2\,ds_{d}$, which probes the parts symmetric in $\cos\theta_d$ in Eq.~(\ref{eq:triple-rate}), we can define an angular asymmetry that probes the contribution linear in $\cos\theta_d$,
\begin{align}\label{eq:d2AFBpi}
\frac{dA_{d}}{ds_d} & = \left(\mathcal{B}_d \frac{d\Gamma}{dq^2}\right)^{-1} \left[\int_0^1 d\cos\theta_d\,d^3\Gamma_d - \int_{-1}^0 d\cos\theta_d\,d^3\Gamma_d \right]\\\nonumber
& = f^d_A(q^2,s_d) A_{\tau}(q^2) + f^d_{\perp}(q^2,s_d)P_{\perp}(q^2).
\end{align}
The energy distribution of this asymmetry is sensitive to both the perpendicular polarization and the tau forward-backward asymmetry. The respective sensitivity of $A_d$ to $A_\tau$ and $P_\perp$ is imprinted in the energy of the decay particle $d$ through the kinematic functions  $f^d_A(q^2,s_d)$ and $f^d_{\perp}(q^2,s_d)$. In Fig.~\ref{fig:properties} we show the differential rate $d\Gamma$ and the decay asymmetries $A_d$ as functions of $q^2$. Integrated over $q^2$, we predict the following asymmetries in the standard model,
\begin{align}
A_\pi = -0.54,\quad A_\rho = -0.32,\quad A_\ell = +0.06.
\end{align}

%---------------------------------------------------------------------------
%---------------------------------------------------------------------------

\section{New physics in polarization observables}\label{sec:np}
\begin{figure}[!tb]
\begin{tabular}{cc}
  \includegraphics[width=64mm]{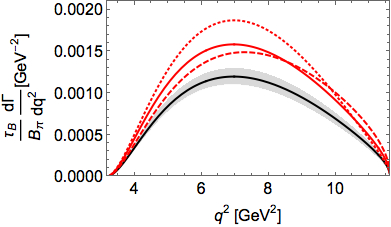}&\hspace{0.5cm}\includegraphics[width=60mm]{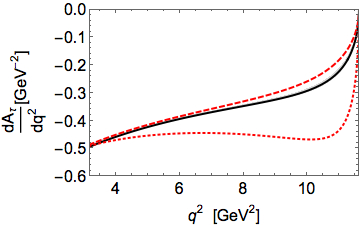}\\  
\hspace{0.25cm}  \includegraphics[width=60mm]{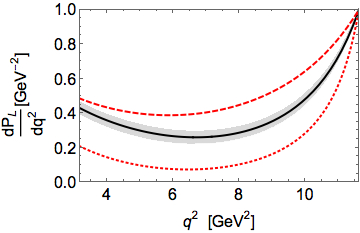}&\hspace{0.45cm}\includegraphics[width=60mm]{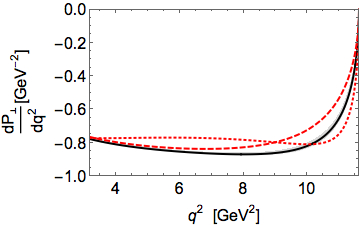}\\
\end{tabular}
\caption{Differential observables of $B\to D\tau^-\bar\nu_\tau$  in the standard model (black) and in scenarios of NP: ``Vector'' (solid red), ``Scalar'' (dashed red) and ``Tensor'' (dotted red).
\label{fig:3body}}
\end{figure}

Measuring additional observables from the kinematic distributions of $B\to D\tau\nu$ is useful to further test the standard model and to
discriminate among the possible New Physics (NP) solutions to the $R_{D^{(\ast)}}$ puzzle. In Fig.~\ref{fig:3body}, we show the $q^2$-spectrum of the total rate, the tau forward-backward asymmetry and the two polarization observables in the standard model (black curves). The uncertainty bands are due to hadronic form factors and correspond to one standard deviation, as implemented in Refs.~\cite{Alonso:2016gym,Alonso:2017ktd}.

We present effects of new physics in three different benchmark scenarios that can explain the $R_{D^{(*)}}$ anomalies. The NP contributions are parametrized in terms of Wilson coefficients corresponding to vector, scalar, and tensor currents~\cite{Alonso:2016gym}. Our benchmarks correspond to
\begin{align}
\text{``Vector''}: & \quad \epsilon_L^\tau=0.15,\\\nonumber
\text{``Scalar''}: & \quad \epsilon^\tau_{S_L}=0.80,\ \epsilon^\tau_{S_R}=-0.65,\\\nonumber
\text{``Tensor''}: & \quad \epsilon^\tau_{T}=0.40.
\end{align}
In the ``Vector'' scenario, only the magnitude of the decay rate is modified by a left-handed vector current, so that effects cancel in the normalized observables $A_\tau$, $P_\perp$ and $P_L$. In the ``Scalar'' and ``Tensor'' scenarios, all differential observables receive sizeable contributions. In Tab.~\ref{tab:results}, we show numerical predictions for the $q^2$-integrated observables in the presence of new physics.
These results demonstrate that the different NP scenarios considered here lead to results unambiguously distinct from the standard model and from each other~\cite{Alonso:2016gym,Alonso:2017ktd,Asadi:2018sym}.  
\begin{table}[!tb]
\centering
\begin{tabular}{|c|c|ccc|}
\hline
&$R_D$&$A_{\tau}$&$P_L$&$P_\perp$\\
\hline
Standard model& $0.299(3)$~\cite{Bigi:2016mdz,Bernlochner:2017jka} &$-0.359(3)$&0.34(3)&$-0.839(7)$\\
\hline
Vector&0.410&$-0.359$&0.34&$-0.839$\\
Scalar&0.400&$-0.335$&0.52&$-0.758$\\
Tensor&0.467&$-0.451$&$0.14$&$-0.779$\\
\hline
Measurement& $0.407(39)(24)$~\cite{HFLAV16} &--&--&--\\
\hline
\end{tabular}
\caption{Numerical predictions of the integrated observables in the standard model and in the different
benchmark scenarios of NP (see main text). \label{tab:results}}
\end{table} 

\section{Observation prospects at $B$ physics experiments}
To estimate the sensitivity to the tau properties at the BELLE II experiment, we perform a statistical analysis, assuming an ideal experiment with unlimited resolution in $q^2$ and $E_d$. Systematic uncertainties are not taken into account. For the longitudinal polarization, the statistical uncertainty $\delta P_L(q^2)$ and the sensitivity $S_L(q^2)$ of a measurement in the differential rate $d^2\Gamma_d/dq^2ds_d$ are given by\begin{align}
\delta P_L(q^2) = \frac{1}{\sqrt{N(q^2)}S_L(q^2)},\qquad S_L^2(q^2) = \int ds_d \frac{f_L(q^2,s_d)^2}{f_0(q^2,s_d) + f_L(q^2,s_d)P_L(q^2)},
\end{align}
In Tab.~\ref{tab:pl}, we present the resulting relative uncertainty on the longitudinal polarization $P_L$, averaged over the momentum range $q^2$. The predictions are made for $3000$ $B^-\to D^0 \tau^-\bar\nu_\tau$ events, as expected with $50\,\text{ab}^{-1}$ of data at BELLE II. We also show predictions for the perpendicular polarization $P_\perp$ and the tau forward-backward asymmetry $A_\tau$, which can be extracted from a combined fit to the energy distribution of the decay asymmetry, $dA_{d}/ds_d$. Correlations between $P_\perp$ and $A_\tau$ are taken into account. It is apparent that $\tau\to\pi\nu$ has the highest sensitivity to the polarizations, while $\tau\to\rho\nu$ competes for $A_\tau$.

%---------------------------------------------------------------------------
\begin{table}[!tb]
\centering
\renewcommand{\arraystretch}{1.5}
\begin{tabular}{|c|c|c|c|}
\hline
$\mathcal{L} = 50\,\text{ab}^{-1}$ & $\tau\to\pi\nu$ & $\tau\to\rho\nu$ & $\tau\to\ell\nu\bar{\nu}$ \\
\hline
$\delta P_L/P_L$ & $0.03$ & $0.07$ & $0.09$ \\
$\delta P_\perp/|P_\perp|$ & $0.09$ & $0.25$ & $0.57$ \\
$\delta A_\tau/|A_\tau|$ & $0.11$ & $0.10$ & $0.40$ \\
\hline
\end{tabular}
\caption{Predicted relative statistical uncertainties on the tau polarizations, $P_L$ and $P_\perp$, and forward-backward asymmetry, $A_\tau$, at BELLE II with $3000$ $B^-\to D^0 \tau^-\bar\nu_\tau$ events.}
\label{tab:pl}
\end{table}
%---------------------------------------------------------------------------

\section{Conclusions}
We have shown that the properties of the tau lepton in $B\to D\tau \nu$ decays can be extracted from the tau's decay products with a good sensitivity. BELLE II has the potential to achieve measurements of the longitudinal and perpendicular polarizations, as well as the tau forward-backward asymmetry, in the standard model at the level of $10\%$. This prediction crucially relies on hadronic tau decays, which have the highest sensitivity to the tau properties. 

Beyond the standard model, tau polarimetry is a strong tool to detect potential new physics contributing to tau production. In the light of the observed discrepancies in the total decay rates of $B\to D \tau \nu$ and $B\to D^\ast \tau\nu$, tau properties are useful to disentangle effects of new physics and probe their viability in different observables. While we have focused on $B\to D\tau\nu$, an extension of our approach to polarization observables in $B\to D^\ast \tau\nu$ is in progress. We encourage experimentalists at BELLE II to exploit the rich opportunities that hadronic tau decays offer.

\section*{Acknowledgements}
We thank the organizers of Tau 2018 for an interesting and enjoyable workshop in Vondelkerk. JMC acknowledges support from the Spanish MINECO through the ``Ram\'on y Cajal'' program RYC-2016-20672. SW acknowledges funding by the Carl Zeiss foundation through an endowed junior professorship (\emph{Junior-Stiftungsprofessur}).

% TODO: include funding information
%\paragraph{Funding information}
%Authors are required to provide funding information, including relevant agencies and grant numbers with linked author's initials. Correctly-provided data will be linked to funders listed in the \href{https://www.crossref.org/services/funder-registry/}{\sf Fundref registry}.

\bibliography{tau2018-westhoff.bib}

\nolinenumbers

\end{document}